\documentclass[aps,twocolumn,pra,floatfix,showpacs,superscriptaddress,10pt]{revtex4-1}

\usepackage[dvips]{graphicx}
\usepackage{latexsym}
\usepackage{amsmath}
\usepackage{amssymb}
\usepackage{amsfonts}
\usepackage{color}
\usepackage{bm}
\usepackage{verbatim}
\usepackage[english]{babel}
\usepackage[utf8]{inputenc}
\usepackage{euscript}
\usepackage{subfigure}
\usepackage[normalem]{ulem}

\usepackage[colorlinks,bookmarks=false,citecolor=blue,linkcolor=blue,urlcolor=blue]{hyperref}
\usepackage[all]{hypcap} 
\begin{document}

\title{Recovery of the electron-phonon interaction function in superconducting tantalum ballistic contacts}

\author{N. L. Bobrov}
\affiliation{B. Verkin Institute for Low Temperature Physics and Engineering, 47, Nauki Ave., 310164 Kharkov, Ukraine\\
Email address: bobrov@ilt.kharkov.ua}
\published {\href{https://fntr.ilt.kharkov.ua/fnt/pdf/45/45-5/f45-0562r.pdf}{Fiz. Nizk. Temp.}, \textbf{45}, 562 (2019); [\href{https://doi.org/10.1063/1.5097356}{Low Temp. Phys.} \textbf{45}, 482(2019)]}
\date{\today}

\begin{abstract}The experimentally observed nonlinearities of the current-voltage characteristics (CVCs) of tantalum-based point homo- and hetero- contacts in both normal and superconducting states related to electron-phonon interaction (EPI)  were  analyzed.  It  was  taken  into account that additional nonlinearity of CVCs arising upon contact transition to the superconducting state (superconducting spectral component) is formed not only near the constriction in the region roughly equal to the contact diameter (as is the case for the normal     state, and as predicted theoretically for the superconducting state), but also in a markedly larger region that is about the size of the  coherence length. In this case, a considerable role in the formation of this superconducting component is played by nonequilibrium  phonons with low group velocity, which account for the experimentally observed sharpening  of the  phonon  peaks  in  the  EPI  spectra (the second derivatives of the CVCs) during the superconducting transition of the contacts, instead of the theoretically expected peak broadening (spreading), and for the increase in the superconducting contribution to the point contact spectrum in the low and medium energy regions.  The high-energy  part of the EPI spectrum changes much less significantly  during  the superconducting  transition, which  is attributable to suppression of the excess contact current by nonequilibrium quasi-particles. A detailed procedure was proposed for the recovery of the EPI spectral function from the point contact spectrum contribution  (the  second  derivative  of  the  CVC)  that  arises during the superconducting transition of one or both contacting metals.


\end{abstract}

\maketitle

\section{INTRODUCTION}
The usage of the nonlinearity of the current-voltage characteristics
(CVC) of ballistic point contacts in the normal state to
restore the electron-phonon interaction (EPI) function in metals
is well known. Several hundred publications are available on this
subject and two integrating monographs have been published \cite{1,2}.
Electron duplication takes place for these contacts in the on-state,
namely, the electrons are split into two groups for which the
energy difference between the occupied and unoccupied states on
the Fermi surface is \emph{eV} \cite{3}. In other words, an electron that has traveled
through the constriction from the opposite contact bank differs in
energy from electrons within the present bank by exactly the applied
voltage value. At any point of the trajectory, the electron may lose
excess energy by emitting a nonequilibrium phonon with the energy
\emph{eV}. Since the contact is ballistic, the average energy relaxation length
is much greater than the contact size.

Thus, the greater part of nonequilibrium phonon emissions
would occur away from the constriction, within the contact banks.
If after scattering, the electron goes back through the constriction to
the initial bank of the point contact from which it flew out with an
additional energy, the contact resistance would increase. These processes
are referred to as backscattering. These processes provide the
basis for the Yanson EPI spectroscopy where the second derivative
of the contact CVC directly reflects the structure of the EPI function.
It is obvious that, since during excess energy electron scattering
via emission of nonequilibrium phonons, in the isotropic case, the
direction of electron movement can change in an arbitrary way,
then, due to geometric considerations, the backscattering processes
are efficient only in a volume relatively proximate to the constriction.
If the contact is shaped like an orifice of diameter \emph{d} in a thin
wall, this volume is approximately equal to the volume of a sphere
with the same diameter \cite{4}. Hence, most of the electron scattering on nonequilibrium phonons does not participate in the formation of
the EPI spectrum in Yanson spectroscopy.

The transition to the superconducting state brings about
an additional CVC nonlinearity caused by the suppression (decrease)
of the excess current (CVC difference between the superconducting
and normal states of the contact metal at the same
voltage) upon reabsorption of nonequilibrium phonons by
Andreev-reflected electrons (hereinafter referred to as Andreev
electrons, for brevity). This nonlinearity can also be used for
the recovery of the EPI function. The appropriate theories that
allow this to be done for S-c-N and S-c-S contacts (S is superconductor,
N is normal metal, c is constriction) appeared in
1983 \cite{5,6}; however, practical attempts to recover the EPI function
from the superconducting characteristics of these contacts were
made relatively recently \cite{7,8,9}.

For tantalum-based contacts, the previously used procedure
for EPI function recovery required further development, taking
account of the ratio between the contact size and the coherence
length. The approaches considered in this article may prove to be
fairly important in the analysis of experimental data of point
contact EPI spectroscopy in superconductors.

\section{THEORETICAL GROUNDS}
Reabsorption processes, in other words, Andreev electron
reflections on nonequilibrium phonons, are not subject to geometrical
constraints typical of backscattering processes; any
reflection process is effective. These processes can take place in
the volume where nonequilibrium phonons and Andreev electrons
coexist simultaneously. Since the conversion of Andreev
electrons to Cooper pairs occurs at the reduced coherence length $\zeta$ \linebreak
 ($\tfrac{1}{\zeta }=\tfrac{1}{{{\xi }_{0}}}+\tfrac{1}{{{l}_{i}}}$, ${\xi }_{0}$
 is the superconducting coherence length, $l_i$  is
 the length of scattering on impurities), then the volume, in the
isotropic case, is equal to the volume of a sphere (or hemisphere
for S-c-N point contacts) with the radius ${\zeta}$.

However, for ballistic contacts with great coherence and elastic
relaxation lengths appearing in the theory, these processes are fairly
probable in approximately the same volume as backscattering. Here
the volume restriction is caused by fast decrease in the concentrations
of both nonequilibrium phonons and Andreev electrons
moving away from the constriction. High current density in the
vicinity of the orifice provides also their high concentration, which
rapidly declines as the current spreads. Therefore, at large distances
\emph{r} from the constriction, the contact can be considered to be a point
source of phonons, with their density decreasing as ${\sim{\ }1}/{{{r}^{2}}}\;$. Since the
minimum size in which the superconducting energy gap can vary
coincides with the length $\zeta \gg {d}$ (\emph{d} is the contact diameter), these
scattering processes do not change the gap in the near-contact
region, and suppression of the excess current is due to a minor
decrease in the quantity of Andreev electrons.

The Khlus and Omel'yanchuk theory \cite{5,6} of inelastic point
contact EPI spectroscopy in the S-c-S and S-c-N point contacts
was developed for ballistic contacts, i.e., for those contacts that
obey the condition $d\ll\zeta$, $\upsilon_F/  \omega_D$ ($\upsilon_F$ is the Fermi velocity,
${{\upsilon}_{F}}/{{\omega }_{D}}\sim{{l}_{\varepsilon}}$, where $l_{\varepsilon}$ is the energy free path at the Debye energy
$\hbar\omega_D$).

For S-c-S contacts, it was found that \cite{5}
\begin{equation}
\label{eq__1}
\frac{d{{I}_{exc}}}{dV}(V)\! \negthinspace=-\frac{64}{3R}\left( \frac{\Delta L}{\hbar \bar{v}} \right)\! \negthinspace{{\left[ {{G}^{N}}(\omega )+\frac{1}{4}{{G}^{S}}(\omega ) \right]}_{\omega ={eV}/{\hbar }\;}}
\end{equation}
Here \emph{R} is the contact resistance, \emph{L} is a function that is fairly
sophisticated for arbitrary arguments, $\bar{v}$ is the velocity of electrons
averaged over the Fermi surface, $G^{N}(\omega)$ is the point contact (PC)
EPI function, which is the same as that of point contacts in the
normal state according to the Kulik-Omel'yanchuk-Shekhter
(KOS) theory \cite{2}, $G^{S}(\omega)$ is the superconducting PC EPI function
differing from $G^{N}(\omega)$ by a form factor, $\Delta$ is the superconducting
energy gap. Furthermore, unlike the normal form factor, which is
responsible for the backscattering contribution to the current, in
the case of the superconducting form factor appearing in $G^{S}(\omega)$, the contribution to the current is made by electron-phonon collisions
associated with Andreev reflection-like processes in the
region of the contact, \emph{i.e.}, by of the conversion of quasi-electron
excitations to quasi-holes. A similar expression was obtained for
the S-c-N contact \cite{6}:

\begin{equation}
\label{eq__2}
\begin{matrix}
  \frac{1}{R(V)}-{{\left( \frac{1}{R(V)} \right)}_{\Delta =0}}= \\
  =-\frac{32}{3R}\times \frac{d\Delta }{h}\times \left[ \frac{1}{v_{F}^{(1)}}\times {{G}_{1}}\left( \omega  \right)+\frac{1}{v_{F}^{(2)}}\times {{G}_{2}}\left( \omega  \right) \right] \\
\end{matrix}
\end{equation}
Here $G_{i}(\omega)$ (\emph{i} = 1, 2)  is the EPI function of metals that form the
point contact.

The relative phonon contribution to the excess current at
$eV\sim {{\omega }_{D}}$ is about ${d\cdot {{\omega }_{D}}}/{{{v}_{F}}}$ , i.e., it is small when the condition
$d\ll {{{v}_{F}}}/{{{\omega }_{D}}}$ holds. This small contribution is very important to
ensuring that equal changes in the nonequilibrium phonon flux
density for different biases at the contact cause equal changes in
the excess current. Meanwhile, when suppression of the excess
current is pronounced, this relationship is violated. For this type
of contacts, recovery of the EPI function would require a correction
of the useful signal amplitude in the region of decreasing
excess current \cite{9}.

For ballistic contacts, the current spreading results in Andreev
electrons being spread over a larger space in which the concentration
of nonequilibrium phonons is low; therefore, the relative
decline of excess current for these contacts is minor. The nonlinear
deviation of the point contact resistance from $R_{0}=R$ at $V=0$ caused
by scattering of nonequilibrium phonons on Andreev electrons is
several times smaller than that caused by backscattering processes.
As a result, if $eV\gg\Delta$, the point contact spectrum changes insignificantly
upon transition to the superconducting state. Khlus \cite{6}
described this transformation of the second derivative of the CVC
for the ballistic S-c-N point contact:
\begin{equation}
\label{eq__3}
\begin{matrix}
\frac{1}{R}\cdot \frac{dR}{dV}=\frac{16ed}{3\pi }\cdot \sum\limits_{i=1,2}{\frac{1}{v_{F}^{(i)}}\cdot \int\limits_{0}^{\infty }{\frac{d\omega }{\Delta }\cdot S\left( \frac{\omega -eV}{\Delta } \right){{G}_{i}}(\omega )}}.
\end{matrix}
\end{equation}
where $G_i(\omega)$ are the EPI functions of the normal and superconducting
metals that form the heterocontact, $S(x)$ is the smearing factor:
\begin{equation}
\label{eq__4}
S(x)=\theta (x-1)\frac{2{{\left( x-\sqrt{{{x}^{2}}-1} \right)}^{2}}}{\sqrt{{{x}^{2}}-1}},
\end{equation}
here $\theta(y)$ is the Heaviside theta function.

Upon the superconducting transition, the spectrum is additionally
smeared and at $T\rightarrow{0}$, the resolution is determined by the
value of $\Delta$.
In view of the relationship between the CVC derivative and
the PC EPI function, it follows from expression (\ref{eq__3}) that
\begin{equation}
\label{eq__5}
\tilde{g}_{pc}^{S}(eV)=\int\limits_{0}^{\infty }{\frac{d\omega }{\Delta }S\left( \frac{\omega -eV}{\Delta } \right)g_{pc}^{N}(\omega )}.
\end{equation}

Thus, with the point contact spectra of the heterocontact in the normal and superconducting states at hand, one can compare the results of calculations and experiments. In addition, as shown by Bobrov \emph{et al}. \cite{8},

\begin{equation}
\label{eq__6}
g_{pc}^{S}(eV)=\frac{1}{\Delta }\int\limits_{0}^{eV}{\left[ \tilde{g}_{pc}^{S}(\omega )-g_{pc}^{N}(\omega ) \right]d\omega }
\end{equation}

Equation (\ref{eq__6}) obviously follows from the fact that the first derivative of the excess current is proportional to the EPI current. It can be seen that the function $\tilde{g}_{pc}^{S}(eV)$ reflects the transformation of the spectrum as the contact goes to the superconducting state. It is proportional to the CVC second derivative, shifted to lower energy by a value of about $\Delta$, additionally broadened and, as a result, it has a slightly lower intensity. The function $g_{pc}^{S}(eV)$ is proportional to the first derivative of excess current and does not
contain additional broadening; according to calculations, its shift to
lower energy is approximately two times less than that of $\tilde{g}_{pc}^{S}(eV)$.

\section{COMPARISON OF THE THEORETICAL CONCLUSIONS
WITH THE EXPERIMENT. TIN-BASED POINT CONTACTS}

\begin{figure}[tbp]
\includegraphics[width=8cm,angle=0]{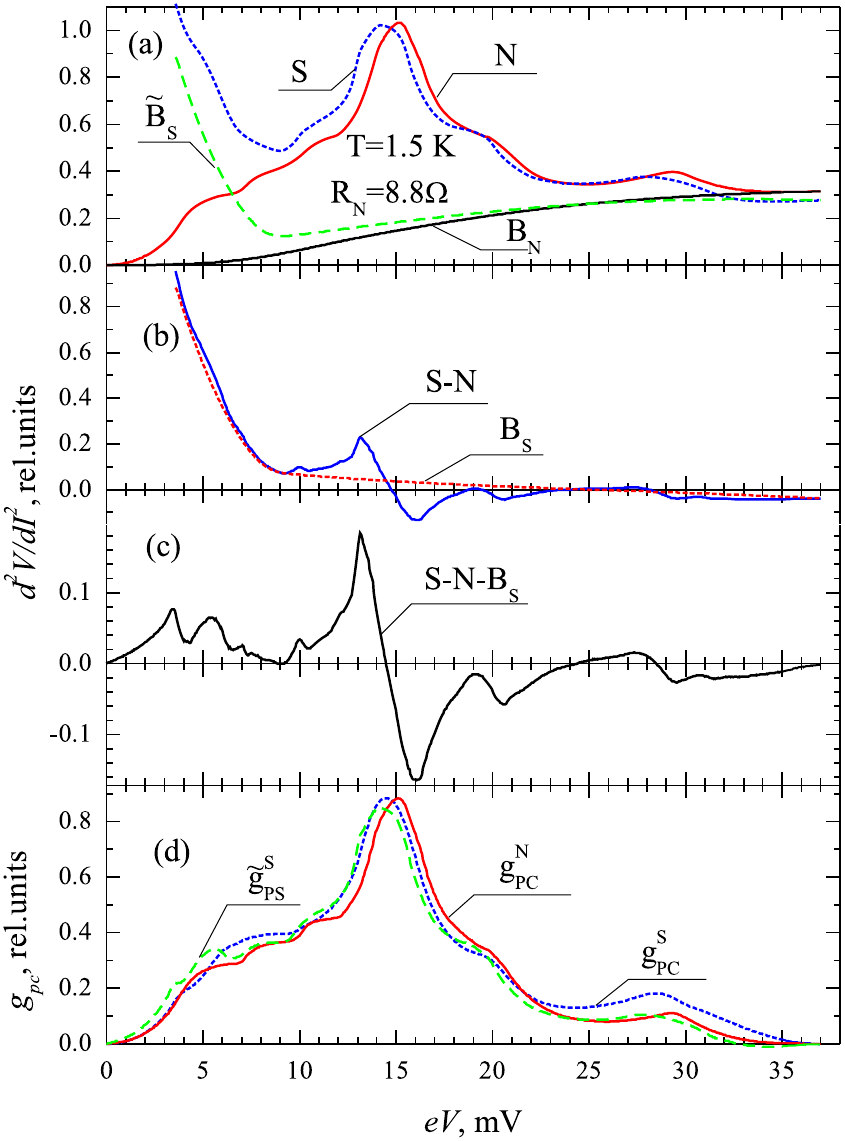}
\caption[]{(a) EPI spectra of the Sn-Cu point contact in the normal and superconducting
states. ${{\tilde{B}}_{S}}$ and $B_N$ are the background curves for the superconducting
and normal spectra, respectively. The superconductivity is suppressed by a
magnetic field. (b) Difference between the superconducting and normal spectra
and the assumed shape of the background curve. (c) The difference curve
(after background subtraction). (d) Point contact EPI functions recovered from
the normal and superconducting states and by integration of the difference
curve. For the convenience of comparison, the curve $g_{pc}^{S}$ is aligned in amplitude with the curve $g_{pc}^{N}$.}
\label{Fig1}
\end{figure}

As an example, first, we will consider tin-based point contacts,
which show good agreement between the theory and the experiment \cite{8}.
 Figure \ref{Fig1}
 presents the spectra of a Sn-Cu point contact in the
normal and superconducting states. The difference curve and the
background curves, $B_N$, $B_S$, and ${{\tilde{B}}_{S}}$, are also given. It turned out that
background is inherent in not only the Yanson point contact spectroscopy.
The background BN is clearly manifested as a nonzero
second derivative of the CVC at displacements greater than those
corresponding to the boundary frequency of the phonon spectrum,
whereas the EPI function goes to zero. The background is determined
most often by the self-consistent iteration procedure \cite{1}.

It was found that mere integration of the difference curve
${{({{d}^{2}}V/d{{I}^{2}})}_{S}}-{{({{d}^{2}}V/d{{I}^{2}})}_{N}}$ is insufficient for obtaining the curve
proportional to the EPI function $g_{pc}^{S}(eV)$, since apart from the effect
related to the gap at low biases, the second derivatives also differ in
the background level beyond the boundary of the phonon spectrum.
This brings about the necessity to subtract the background curve $B_S$
from the difference curve before integration. Apart from the obvious
condition of being equal to zero at energies exceeding the maximum
phonon spectrum frequency, the curve obtained after background
subtraction should obey the \emph{rule of sums: the total areas under the
curves above and below the abscissa axis must be equal}.

The rule of sums follows from the fact that integration of the
obtained curve gives a curve with a zero background. It is clear that
these criteria can be met for a multitude of different background
curves; however, their variations, provided they are monotonic, do
not induce considerable changes in the shapes and positions of
phonon effects of the EPI function being recovered.

In order to verify the theory predictions, it is necessary to
follow the transformation of the spectrum upon transition from
the normal to superconducting state without the background component
${{\tilde{B}}_{S}}$, in order to get a curve proportional to $\tilde{g}_{pc}^{S}(\omega )$. In the
absence of a normal state spectrum and difference curve with subtracted
backgrounds, this is difficult to perform; however, when these
curves are available, this problem is easily solved: let $\tilde{S}$ be the spectrum
in the superconducting state with background subtracted.
Then \linebreak
 ${{\tilde{B}}_{S}}=S-\tilde{S}=S-[(N-{{B}_{N}})+(S-N-{{B}_{S}})]={{B}_{N}}+{{B}_{S}}$. The
latter is even somewhat more convenient, as this allows for an easier
approximation of the missing part of the curve in the small bias
region near the gap effect. For the convenience of comparison, the
curve $g_{pc}^{S}$ in Fig.\hyperref[Fig1]{1(d)} is aligned in amplitude with the curve $g_{pc}^{N}$. As
can be seen from comparison of Figs.\hyperref[Fig1]{1(a)} and \hyperref[Fig1]{1(d)}, the theory predictions
are in good agreement with the experiment.

Upon transition to the superconducting state, the $\tilde{g}_{pc}^{S}(\omega )$ curves
smear, decrease in the amplitude, and shift to lower energies by a
value approximately equal to the gap. The shapes of experimental
and theoretical ${g}_{pc}^{S}$ curves differ in the high-energy region: the experimental
curve is markedly more intense. Apparently, this is attributable
to increasing concentration of nonequilibrium phonons at the
contact periphery caused by the decrease in the energy relaxation
length of electrons close to the Debye energy.

\section{COMPARISON OF THE CHARACTERISTIC
PARAMETERS FOR TIN AND TANTALUM}
It turns out that even the complete correspondence of the point
contact parameters to the requirements of the theory does not ensure
that the experimental curves follow the theoretical model. For
example, as it was ascertained for ballistic point contacts based on
tantalum \cite{10,11,12}, the superconducting transition is accompanied by a
radical change in the shape, intensity, and positions of phonon
effects. Furthermore, differences from theory predictions are observed
for both hetero- and homocontacts. Instead of the expected broadening
of the phonon peaks, they become strongly sharpened. As a consequence,
the amplitude of these peaks substantially increases.

In addition, the soft phonon mode, which is manifested in the
normal state as a shoulder, is converted to a sharp peak as the contacts
switch to the superconducting state. This is manifested most
vividly for relatively low-resistance point contacts, but even for
high-resistance contacts, in which the ballistic condition is satisfied
rather strictly, the deviations from theory predictions are quite
large. In order to understand what could be responsible for these
crucial differences in the behaviors of tantalum- and tin-based
point contacts, compare their characteristic parameters summarized
in \hyperref[Table 1]{Table I}.
\begin{table*}[tbp]
\caption[]{Characteristic parameters of tantalum- and tin-based point contacts.}
\begin{tabular}{|c||c|c|c|c|c|c|c|c|c|c|}\hline
       &\ \ R, Ohm\ \   & \ \ d, nm \ \    & \ \ ${\rho}_{300}/{\rho}_{res} $ \ \ &  \ \ $\rho l, Ohm \cdot cm^2$ \ \ & \ \ $\upsilon_{F}, cm/s$\ \ &\ \  $l_{\varepsilon}^D, nm$ \ \ &\ \  $l_{i}, nm$ \ \ &\ \ $\xi_{0}, nm$ \ \ &\ \  $\zeta, nm$\ \ &\ \  $\Delta, mV$ \\ \hline \hline
Sn     & $7\div30$   & $5.1\div10.5$ & $\sim$15000             & $4.5\cdot10^{-12}$        & $1.89\cdot 10^8$    & $\sim$360               & 6000          & $\sim$200     & $\sim$200   & 0.57 \\ \hline
Ta     & $15\div210$ & $2.2\div8.5$  & 18                      & $5.9\cdot10^{-12}$        & $0.74\cdot 10^8$    & $\sim$90                & 82            & 92            & 43          & 0.71 \\ \hline
\end{tabular}
\label{Table 1}
\end{table*}

\begin{figure}[tbp]
\includegraphics[width=8cm,angle=0]{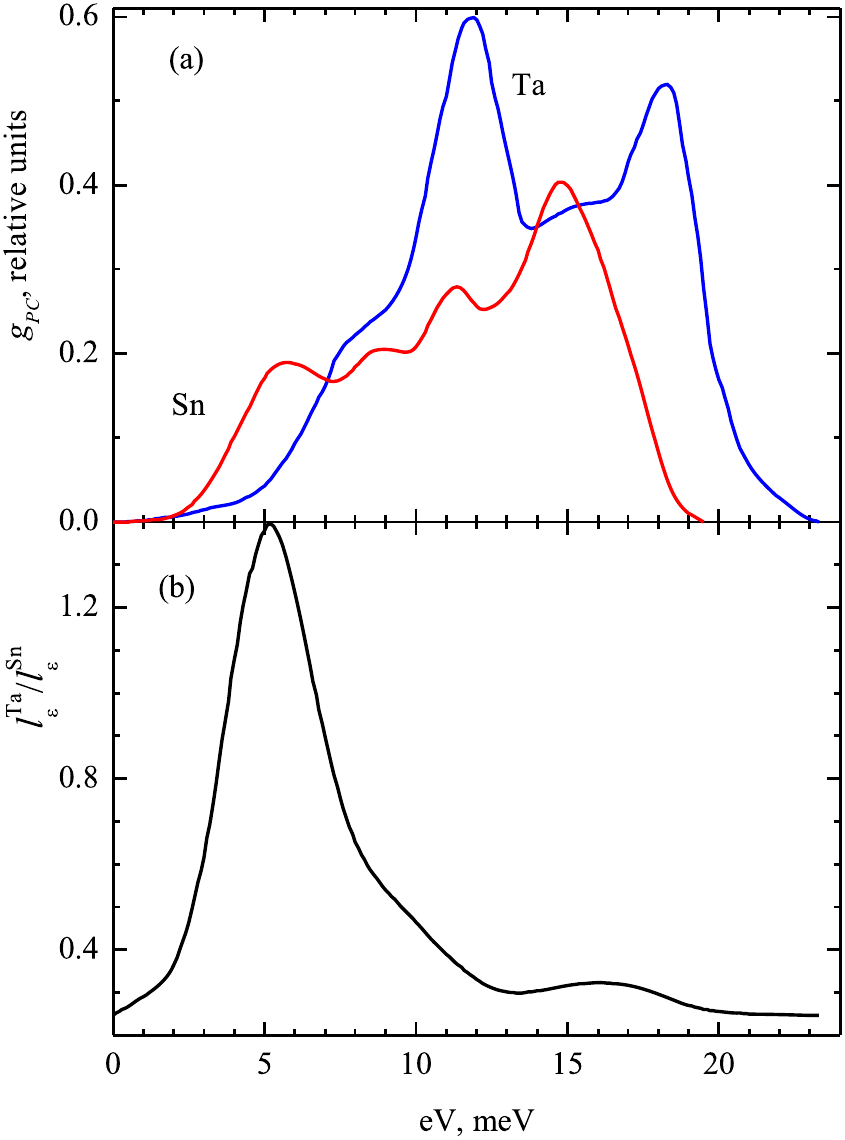}
\caption[]{(a) Point contact EPI functions for tin and tantalum used to estimate the
energy relaxation rate vs. the contact bias. (b) Energy relaxation lengths for tantalum
in relation to tin vs. the bias.}
\label{Fig2}
\end{figure}

Since the Table presents the energy relaxation length at the
Debye energies, $l_{\varepsilon }^{D}$ , and the deviations in the tantalum spectra from
the theory predictions already start at small biases, it appears reasonable
to compare these lengths over the whole range of biases.
The inelastic mean free path of electrons for an arbitrary contact
voltage can be estimated from the formula \linebreak
 $\frac{1}{{{l}_{\varepsilon }}\left( eV \right)}=\frac{2\pi }{\hbar {{v}_{F}}}\int\limits_{0}^{eV}{d\omega \,g(\omega )},$
where $g(\omega)$ is the EPI function.  Figure \ref{Fig2}
shows the point contact
EPI functions for Ta \cite{13} and Sn \cite{1} that were used to estimate the
energy relaxation lengths. The estimates were made using the electron
parameters presented in the Table. As can be seen in Fig.\ref{Fig2},
the relaxation length is greater in tantalum than in tin and is in the
4-7~meV range, which is close to the soft mode energy in tantalum.
At other energies, $l_\varepsilon$ is smaller in tantalum than in tin; it is approximately
four times smaller at the Debye energies.

Considering these estimates, one can conclude that the
reduced coherence length $\zeta$ is the key factor responsible for the
deviation from the theory`s predictions. As follows from the \hyperref[Table 1]{Table},
this value for tin coincides with the coherence length, $\zeta\approx\xi_0$, in
view of the very large elastic scattering length $l_i$, while in tantalum,
$\zeta\sim$~43~nm, i.e., the volume of the sphere limited by the reduced
coherence length is 100 times smaller in tantalum than in tin. For
these reasons, and because of the lower Fermi velocity of electrons
in tantalum, the concentration of Andreev electrons in this volume
increases compared with that for tin. Therefore, reabsorption of
nonequilibrium phonons by Andreev electrons starts to play a
noticeable role not only in the volume roughly equal to the contact
diameter, as in tin, but also in the volume with the characteristic
size of the reduced coherence length. In other words, the excess
current decreases not only because of decreasing quantity of the
Andreev electrons within a volume about equal to the contact
diameter, but also because of suppression of the gap in the contact
region with a volume about equal to the coherence length.

\section{NONEQUILIBRIUM EFFECTS IN TANTALUM-BASED
POINT CONTACTS}

Before considering specific features of the formation mechanism
of phonon peaks in the tantalum-based superconducting contacts,
we will pay attention to other nonlinear effects manifested in
the CVCs of these contacts, which are also associated with the
small coherence length. These effects are not spectral and their
position in the energy axis depends on both the point contact resistance
and the temperature and magnetic field strength. In the
second derivative of the CVC, the low-temperature effect occurs
immediately after the gap effect as a narrow and sharp surge and is
greater in intensity. This corresponds to a stepwise decrease in the
excess current caused by the abrupt decrease in the superconducting
gap in the region adjoining the contact. This phenomenon is
associated with the attained critical density of nonequilibrium
quasi-particles in the near-contact region and is considered in
detail in Refs.\cite{10,11,12}. Since the contacts investigated here are ballistic,
most of electrons lose excess energy within the point contact
banks, when scattered on nonequilibrium phonons.

In the superconducting state these electrons that have lost the
excess energy are accumulated above the gap. With increasing contact
bias, the quantity of these nonequilibrium quasi-particles increases;
when some critical concentration is reached, a part of the superconductor
adjoining the constriction abruptly switches to a new, nonequilibrium
state with a partly suppressed gap. It is obvious that the
smaller the space in which this switching takes place, the more
rapidly the required concentration is reached. The minimum size of
this region cannot be smaller than the reduced coherence length $\zeta$.
The position of the nonequilibrium effect at a specified temperature
for contacts with different resistance corresponds to the same injection
power, being approximately 0.4~$\mu$W at 2K.

As the temperature or magnetic field increases, the relaxation
rate of these quasi-particles above the gap increases; therefore, for
attaining the critical concentration, it is necessary to increase the injection
power. This shifts the effect to higher voltages. This rules out the
interpretation of these effects as being due to the degradation of superconductivity
caused by heating or suppression by the magnetic field.
Since the stepwise decrease in the excess current upon transition to
the nonequilibrium state is insignificant, the change in the gap accompanied
by the transition is also moderate. Thus, the influence of the
nonequilibrium effect on the superconducting spectral contribution is
also small, except that it can be located at biases corresponding to
some phonon effect and thus hamper observation of the phonon
effect. However, as follows from experimental results, the suppression
of the excess current is not limited to the stepwise section resulting
from the phase transition to a new nonequilibrium state.

On further increase of the voltage applied to the contact and,
hence, increase in the injection power, the suppression of the excess
current is rather smooth and is not accompanied by any indications
in the CVC derivatives. Depending on the resistance, this
suppression may be rather small for high-resistance contacts and
fairly large (several-fold larger) for low-resistance contacts. In any
case, this suppression has a pronounced influence on the generation
of EPI-related effects.

\section{EFFECT OF THE NEAR-CONTACT REGION ON THE
FORMATION OF THE SPECTRUM}

Now we will consider the mechanism of formation of phonon
effects related to the excess current. Conventionally speaking, they
represent a superposition of contributions from two spatially different
regions: a contribution from the region about equal to point contact
diameter, which corresponds to the theoretical model, and a contribution
of the near-contact region about equal to the coherence length
$\zeta$. In principle, the mechanism of formation of phonon effects is the
same in both regions and is related to reabsorption of nonequilibrium
phonons by Andreev electrons. However, since the second region is
much larger than the contact diameter, here a considerable role is
played by differences between the group velocities of nonequilibrium
phonons generated by electrons with the excess energy $eV$.

Since phonons with energies corresponding to the maximum
density of states have the lowest group velocities, $\partial \omega /\partial q=0$, these
phonons would leave this region more slowly and would thus be
accumulated in a higher concentration. Since the relative concentrations
of nonequilibrium phonons and Andreev electrons determine
the value of the negative contribution to the excess current, the
greatest contribution to the spectrum will occur at the maximum
density of phonon states. This assumption is supported by sharpening
of the phonon peaks. In the region approximately equal to the
contact diameter, because of small volume, slow phonons do not
have time to concentrate, and their specific contribution to the nonlinearity
would be the same as that of phonons with higher group
velocity. In terms of this model, it is clear that for selection of
phonons with low group velocities, fast phonons must be free to
leave the near-contact volume, \emph{i.e.}, the flight of phonons should correspond
to the ballistic regime.

Thus, the ratio of contributions of the near-contact and remote
regions would influence the degree of sharpening of phonon effects.
Evidently, the increase in the contact size would lead to increasing
proportion of the latter contribution and, hence, to increasing degree
of sharpening of phonon effects. As shown below, this assumption is
perfectly confirmed by experimental results.

\section{TA-CU HETEROCONTACTS}

\begin{figure}[tbp]
\includegraphics[width=8cm,angle=0]{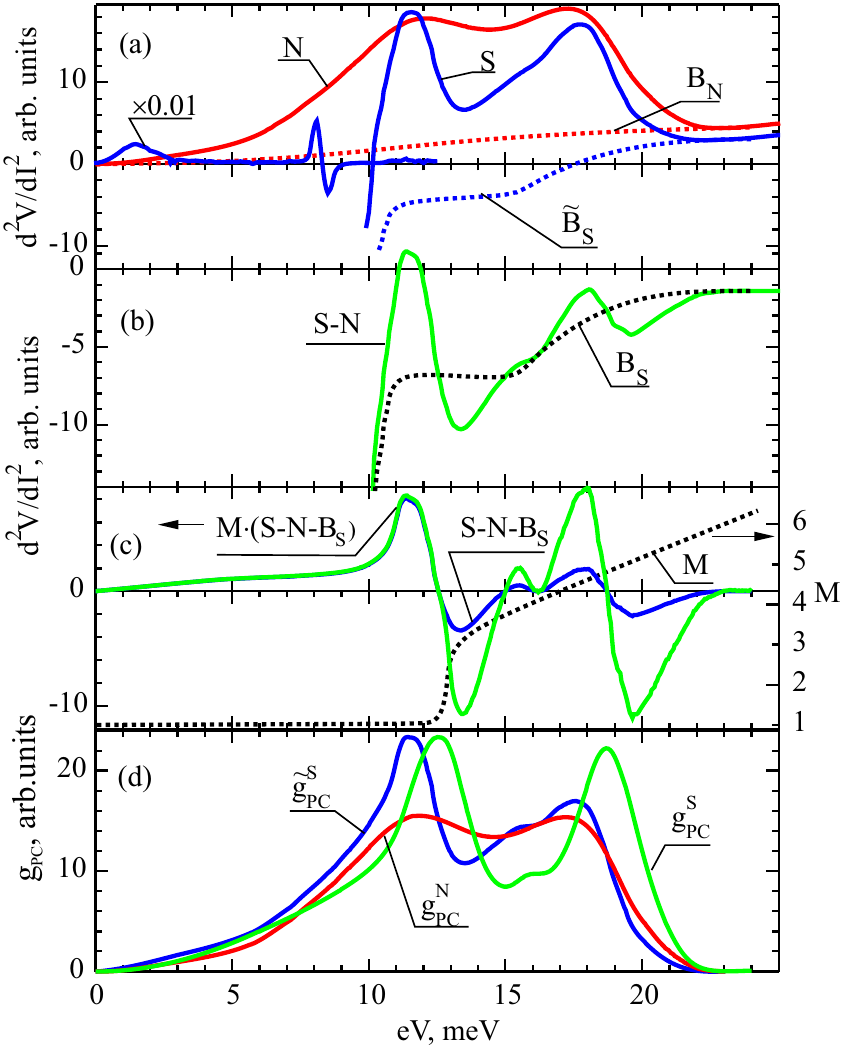}
\caption[]{(a) EPI spectra of the Ta-Cu point contact in the normal and superconducting
states. $T=1.4~K$, $R_0$=210~ohm; the initial segment of the superconducting
curve containing the gap and nonequilibrium effects is scaled down by a
factor of 100, ${{\tilde{B}}_{S}}$ and $B_N$ are the background curves for the superconducting
and normal spectra, respectively. (b) The difference between the superconducting
and normal spectra, and the assumed shape of the background curve. (c)
Difference curve (after background subtraction), scaling curve M, and the difference
curve multiplied by the scaling curve. (d) Point contact EPI functions
recovered from the spectra for normal and superconducting states and from the
superconducting contribution to the spectrum by integration of the corrected difference
curve (c). For convenience of comparison, the curve $g_{pc}^{S}$ is aligned in
amplitude with the curve $\tilde{g}_{pc}^{S}$. The scale is the same in all panels of the Figure.}
\label{Fig3}
\end{figure}

Unlike the spectra of Sn-Cu heterocontacts, the spectra of
Ta-Cu heterocontacts do not show the contribution of copper \cite{13}.
In the spectra of heterocontacts of transition \emph{d}-metals with Cu, Ag,
and Au, the contribution of the latter is not manifested. To begin
with, consider a high-resistance (209~ohm) Ta-Cu heterocontact (see
Fig.\ref{Fig3}).
Formally, it satisfies all requirements of the theory: its diameter
is $\approx$2.2~nm, whereas the elastic scattering length in the tantalum
bank of the contact at the liquid helium temperature is $\approx$84 nm, the
energy relaxation length at the Debye energy is approximately the
same ($\approx$90 nm), and the reduced coherence length in tantalum is
$\zeta\approx$43~nm; in other words, all of the lengths are more than an order
of magnitude greater than the contact diameter. Nevertheless, the
transformation of the spectrum upon transition to the superconducting
state differs crucially from predictions of the theory for ballistic
S-c-N point contacts.

Pay attention to the region of phonon energies. First, in the
superconducting state, there is no spectral smearing and no related
decrease in the intensity. Conversely, the phonon modes are markedly
sharpened and their amplitude increases. Since the resistance
of this contact is very high, the above-noted nonequilibrium effect
is located in a rather high-energy region and coincides with the soft
phonon mode; therefore, it is impossible to follow the transformation
of this mode upon the superconducting transition. Regarding
higher frequency regions of the spectrum, they can be examined
most conveniently in Fig.\hyperref[Fig3]{3(d)}, which shows the curves after background
subtraction. A comparison of $g^N_{pc}$ and $\tilde{g}_{PC}^{S}$ demonstrates that
upon transition to the superconducting state, the amplitude of the
first phonon peak increases to the highest extent, while the growth
of the high-energy peak is much less pronounced. Therefore, it is
impossible to perform a correct recovery of the EPI function from
the superconducting contribution to the spectrum, after the same
background subtraction procedure as was done for tin, by integration
of the S-N-$B_S$ curve, because the rule of sums is not satisfied
in this case. This inconsistency was resolved by making a correction
of the shape of the S-N-$B_S$ curve, which was done by multiplying
it by the scaling curve M [see Fig.\hyperref[Fig3]{3(c)}]. This empirical curve does
not change the difference curve in the low-frequency region and
increases its amplitude in the high-frequency region.

\begin{figure}[tbp]
\includegraphics[width=8cm,angle=0]{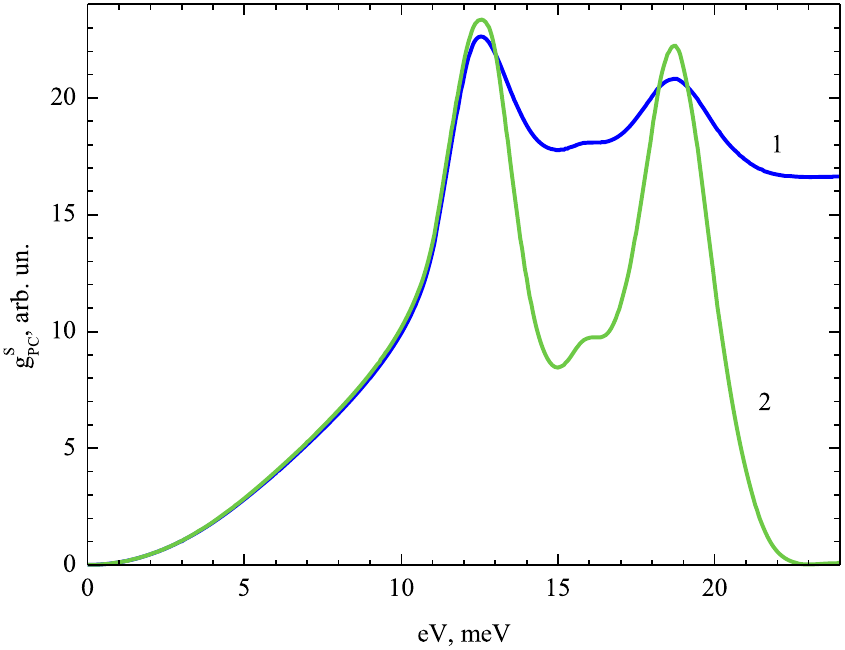}
\caption[]{A demonstration of how violating the rule of sums impacts the shape of
the recovered EPI function. EPI function with the background (1) recovered
without accounting for the correction of the suppressed high-frequency part of
the S-N-$B_S$ curve [see Fig.\hyperref[Fig3]{3(c)}] and without the background (2) recovered
from the corrected curve M(S-N-$B_S$). The scale of the curves is the same as
that in Fig.\ref{Fig3}.}
\label{Fig4}
\end{figure}

The result of recovery before and after the correction can be
observed in Fig.\ref{Fig4}.
The decrease in the amplitude of the superconducting
contribution to the spectrum in the high-energy region is caused
by suppression of the excess current in the contact by nonequilibrium
quasi-particles mentioned above in the discussion dealing with nonspectral
nonequilibrium effects in superconducting curves.
\begin{figure}[tbp]
\includegraphics[width=8cm,angle=0]{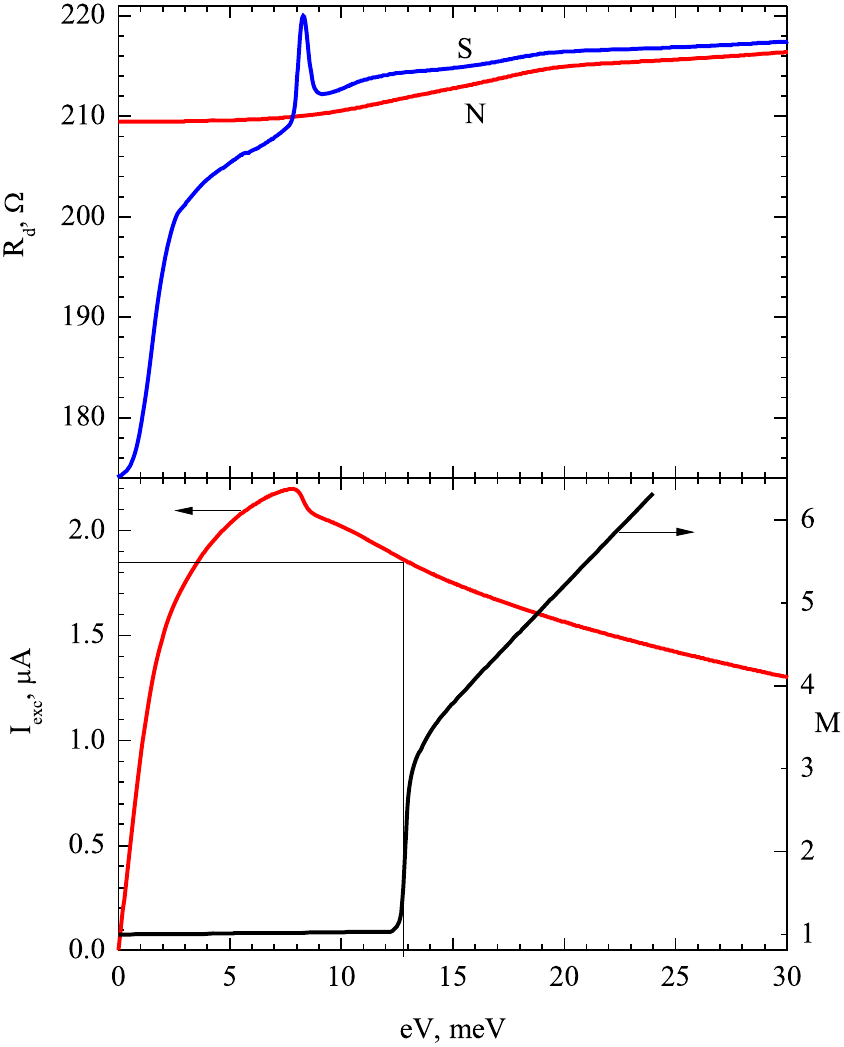}
\caption[]{Differential resistance of the point contact whose characteristics are
shown in Fig.\ref{Fig3} in the normal (N) and superconducting (S) states and excess
current vs. the bias. The excess current panel shows the scaling curve M (see
Fig.\ref{Fig3}) with designated excess current and voltage values corresponding to the
step position in the scaling curve.}
\label{Fig5}
\end{figure}
Figure \ref{Fig5} presents the differential resistances of the point
contact in the normal and superconducting states, the dependence
of the excess current on the voltage, and the scaling curve for correction
of the amplitude of the superconducting contribution to
the spectrum [the same as in Fig.\hyperref[Fig3]{3(c)}]. The vertical and horizontal
segments refer to the voltage and excess current corresponding to
the onset of suppression of the amplitude of the superconducting
contribution to the spectrum.
\begin{figure}[tbp]
\includegraphics[width=8cm,angle=0]{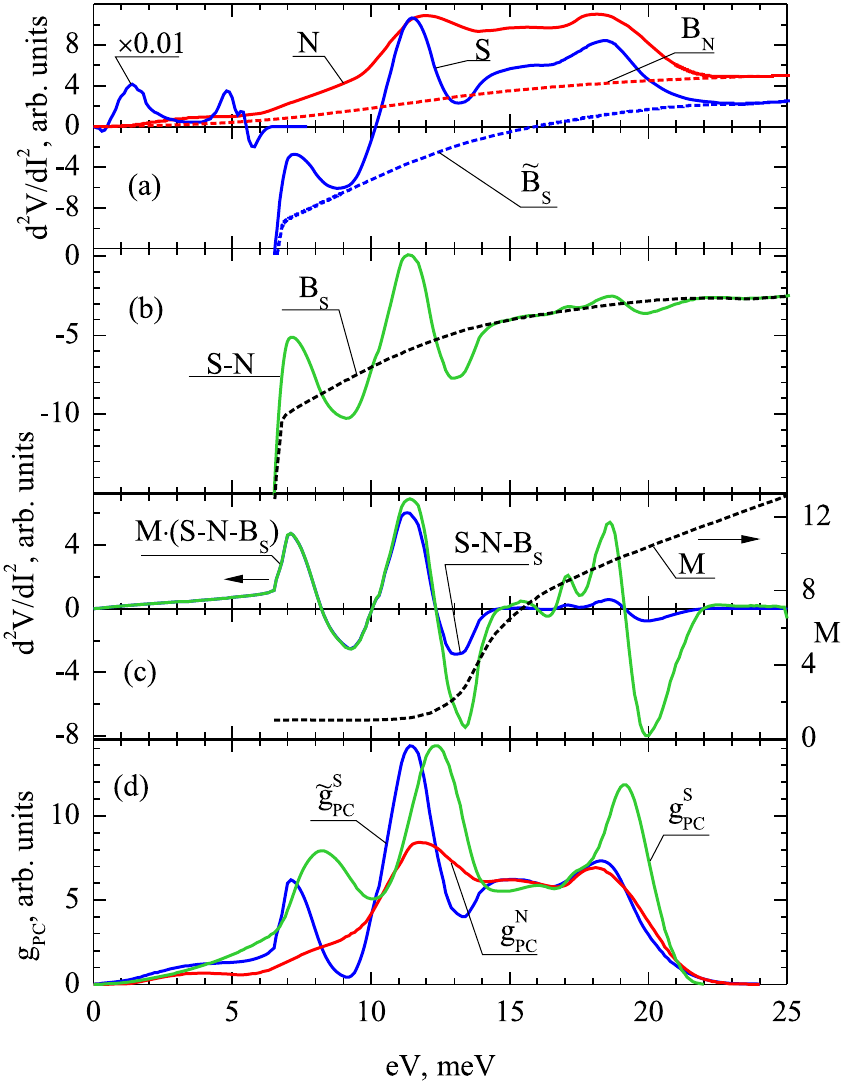}
\caption[]{(a) EPI spectra of the Ta-Cu point contact in the normal and superconducting
states. $T=1.7 K$, $R_0$=73~ohm; the initial section of the superconducting
curve containing the gap and nonequilibrium effects is scaled down by a factor of
100, ${{\tilde{B}}_{S}}$ and $B_N$ are the background curves for the superconducting and normal
spectra, respectively. (b) The difference between the superconducting and
normal spectra and the assumed shape of the background curve. (c) Difference
curve (after background subtraction), scaling curve M, and the difference curve
multiplied by the scaling curve. (d) Point contact EPI functions recovered from
the spectra for normal and superconducting states and from the superconducting
contribution to the spectrum by integration of the corrected difference curve (c).
For the convenience of comparison, the curve $g_{pc}^{S}$ is aligned in amplitude with
the curve $\tilde{g}_{pc}^{S}$. The scale is the same in all panels of the Figure.}
\label{Fig6}
\end{figure}

Let us consider the influence of the decrease in the point
contact resistance. Figure \ref{Fig6} shows the spectra of the point contact
with 73~ohm resistance. In this case, the nonequilibrium effect
already occurs at about $\sim$5mV and coincides with the energy of the
tantalum soft phonon mode by only an edge. The shape of this
mode in the superconducting state basically differs from that in the
normal state and is manifested as a peak rather than a shoulder; the
phonon peak at about 11.3~mV is also markedly sharper than that in
the case of the 209~ohm contact. If one refers to Fig.\hyperref[Fig6]{6(d)} , in the lowenergy
part of the spectrum, attention is attracted by a much more
pronounced increase in the peak amplitude upon transition to the
superconducting state in comparison with the previous point
contact; this supports the above idea that the contribution of the
peripheral areas to the whole superconducting contribution increases
with increasing point contact diameter. However, the superconducting
contribution for the high-energy part of the spectrum is markedly
lower in this case than for the previous contact.
\begin{figure}[tbp]
\includegraphics[width=8cm,angle=0]{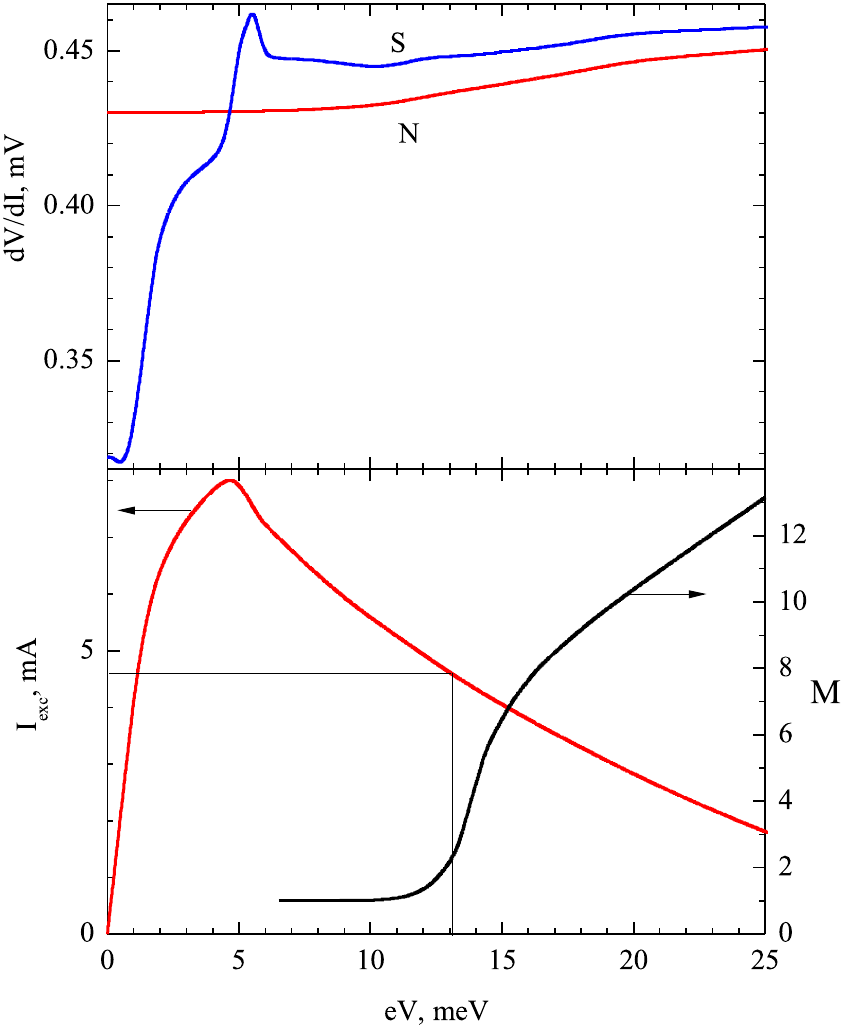}
\caption[]{The first derivatives of the CVCs for a point contact whose characteristics
are shown in Fig.\ref{Fig6} in the normal (N) and superconducting (S) states and
excess current vs. the bias. The excess current panel shows the scaling curve
M (see Fig.\ref{Fig6}) with designated excess current and voltage values corresponding
to the start of the ascending part of the scaling curve.}
\label{Fig7}
\end{figure}
According to
Fig.\ref{Fig7}, the suppression of the excess current with increasing contact
bias is much greater in this case than in the former case, which is
responsible for the observed result. Owing to these factors together,
the scaling curve for correction of the high-frequency part of the
superconducting contribution is approximately twice greater for this
contact than that for the contact considered above.

\begin{figure}[tbp]
\includegraphics[width=8cm,angle=0]{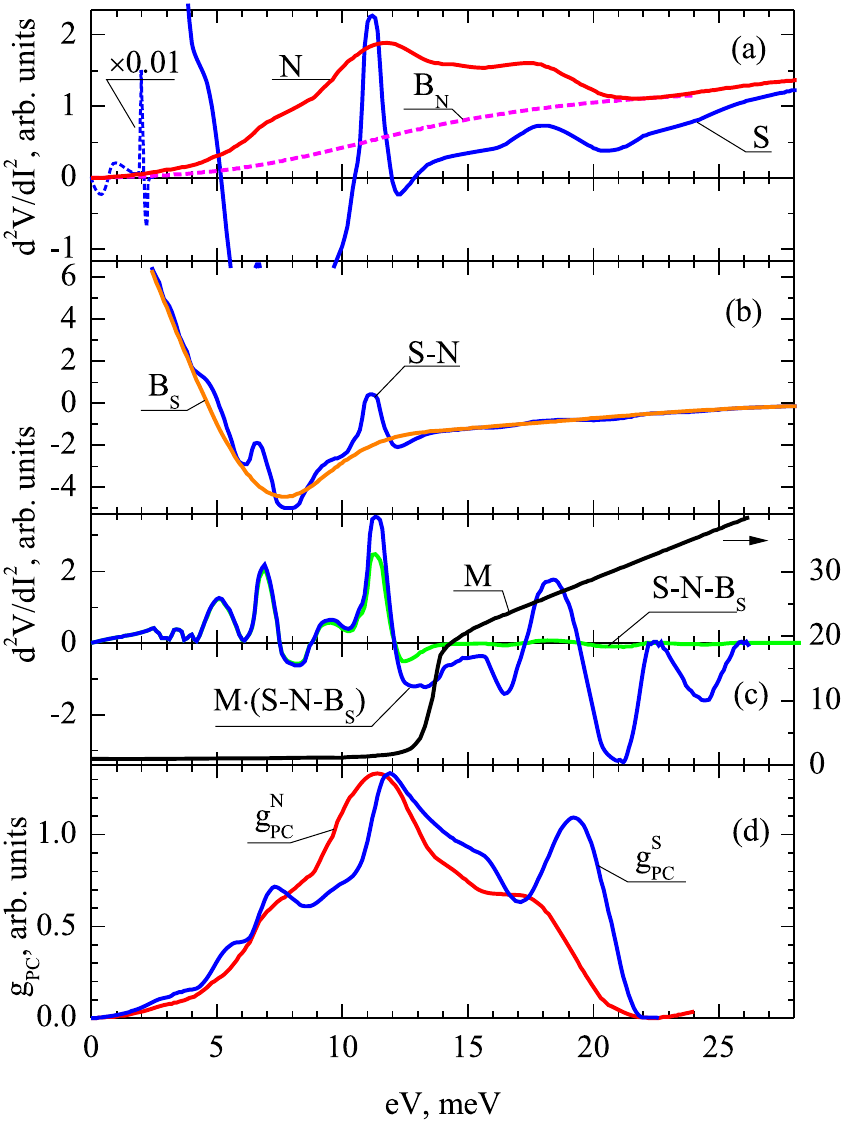}
\caption[]{(a) EPI spectra of the Ta-Cu point contact in the normal and superconducting
states; N: $T=4.6~K$; S: $T=1.7~K$, $R_0$=16~ohm, the initial dashed
segment of the superconducting curve containing the gap and the nonequilibrium
effects is scaled down by a factor of 100, $B_N$ is the background curve for
the normal spectrum. (b) The difference between the superconducting and
normal spectra and the assumed shape of the background curve. (c) Difference
curve (after background subtraction), scaling curve M, and the difference curve
multiplied by the scaling curve. (d) Point contact EPI functions recovered from
the spectra in the normal state and from the superconducting contribution to the
spectrum by integrating the corrected difference curve (c). For the convenience
of comparison, the curve $g^S_{pc}$ is aligned in amplitude with the curve $g^N_{pc}$ . The
scale is the same in all panels of the Figure.}
\label{Fig8}
\end{figure}
Figure \ref{Fig8} shows the EPI spectra for the Ta-Cu point contact
(one of the lowest-resistance point contacts involved in the study).
When its resistance is 16 ohm, the diameter is close to 8.5~nm,
which is almost 4 times greater than that of the previous point
contact (the volume is almost 60 times greater). Here the differences
from the first two point contacts are much more pronounced. For
this contact, the nonequilibrium effect is located almost immediately
after the gap effect at a voltage slightly above 2~mV. This opens up
the possibility to follow the fine structure of the initial section of the
phonon spectrum of tantalum upon the transition to the superconducting
state, because in the normal state, it is possible to follow
only a gradual ascent of the spectrum in this energy range.

\begin{figure}[tbp]
\includegraphics[width=8cm,angle=0]{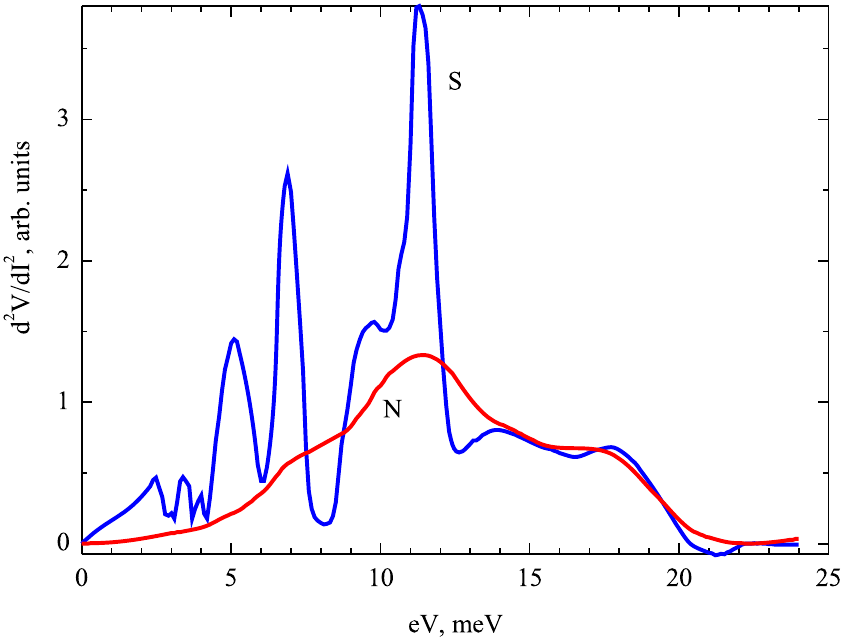}
\caption[]{Second derivatives of the CVCs of the 16~ohm Ta-Cu point contact
(see Fig.\ref{Fig8}) after background subtraction in the normal (N) and superconducting
(S) states. The scale of the curves is the same as that in Fig.\ref{Fig8}.}
\label{Fig9}
\end{figure}
Figure \ref{Fig9} depicts the relevant EPI spectra in the superconducting
and normal states after background subtraction. For this contact in
the superconducting state, the peak sharpening in the low-frequency
region is much more pronounced than in the previous cases. The
spectral amplitude also greatly increases; meanwhile these changes in
the high-energy part of the spectrum are much weaker. In the difference
curve [Fig.\hyperref[Fig8]{8(c)}], the high-energy part is barely seen against the
initial part before the correction.

\begin{figure}[tbp]
\includegraphics[width=8cm,angle=0]{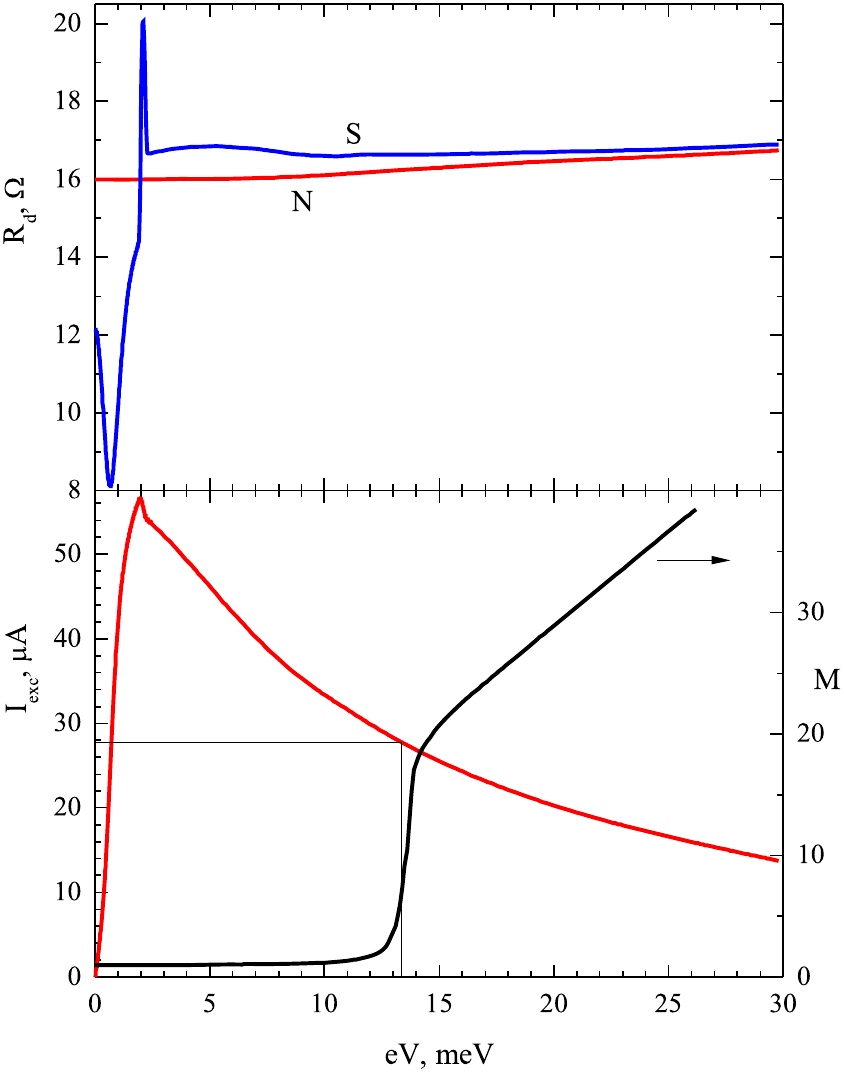}
\caption[]{Differential resistance of the Ta-Cu point contact whose characteristics
are shown in Fig.\ref{Fig8} in the normal (N) and superconducting (S) states and
excess current vs. the bias. The excess current panel shows the scaling curve
M (see Fig.\ref{Fig8}) with designated excess current and voltage values corresponding
to the start of the ascending part of the scaling curve.}
\label{Fig10}
\end{figure}
Figure \ref{Fig10} shows the plots for differential resistances of this
contact in the normal and superconducting states, excess current as
a function of voltage, and the correction scaling curve. As follows
from the Figure, the scaling curve amplitude is approximately 6
times higher than that of the previous contact, which is evidently
attributable to two factors mentioned above: a pronounced increase
in the amplitude of the low-energy part of the superconducting
contribution, and a more pronounced suppression of the highenergy
part caused by decreasing excess current.

\section{Ta-Ta HOMOCONTACTS}
\begin{figure}[tbp]
\includegraphics[width=8cm,angle=0]{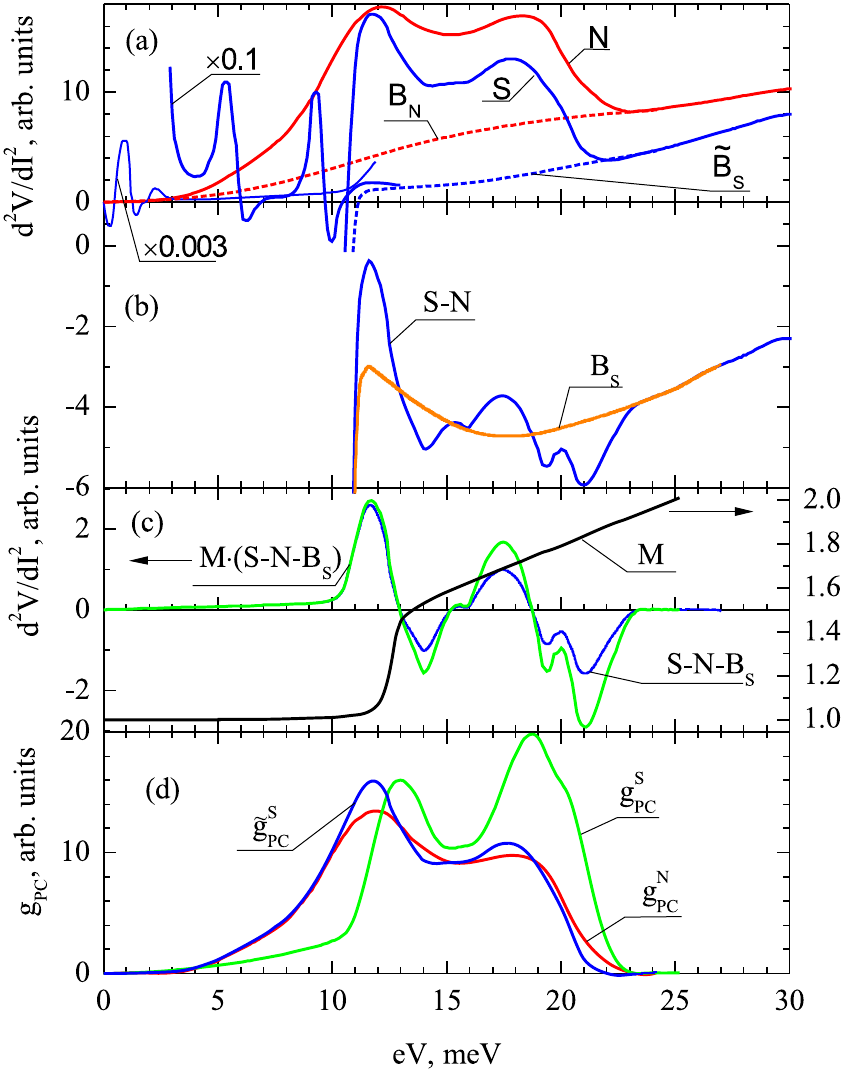}
\caption[]{(a) EPI spectra of the Ta-Ta point contact in the normal and superconducting
states. N: $T=4.6~$K; S: $T=2.0~K$, $R_0$=64~ohm, the initial segments of
the superconducting curve containing the gap and nonequilibrium effects are
scaled down by factors of 300 and 10, respectively; $\tilde{B}_{S}$ and $B_N$ are the background
curves for the normal and superconducting spectra, respectively. (b) The
difference between the superconducting and normal spectra and the assumed
shape of the background curve. (c) Difference curve (after background subtraction),
scaling curve M, and the difference curve multiplied by the scaling curve.
(d) Point contact EPI functions recovered from the spectra for normal and superconducting
states and from the superconducting contribution to the spectrum by
integration of the corrected difference curve (c). The scale is the same in all
panels of the Figure.}
\label{Fig11}
\end{figure}

The trends observed for heterocontacts are generally reproduced
for homocontacts, although differences are also present: the
spectra of homocontacts usually exhibit two nonequilibrium effects
associated with attainment of the critical density of nonequilibrium
quasi-particles in each bank. Figure \ref{Fig11}
 presents the EPI spectra of
the 64~ohm Ta homocontact in the normal and superconducting
states. Since the temperature during the measurements in the
superconducting state proved to be somewhat higher for this
contact than for heterocontacts, and because of the presence of two
nonequilibrium effects, all low-energy region down to less than 10~meV energies was inaccessible for observation of the behavior of
the phonon effects. Therefore, the initial section of the EPI function
recovered from the superconducting contribution to the spectrum
was approximated by a parabola.
\begin{figure}[tbp]
\includegraphics[width=8cm,angle=0]{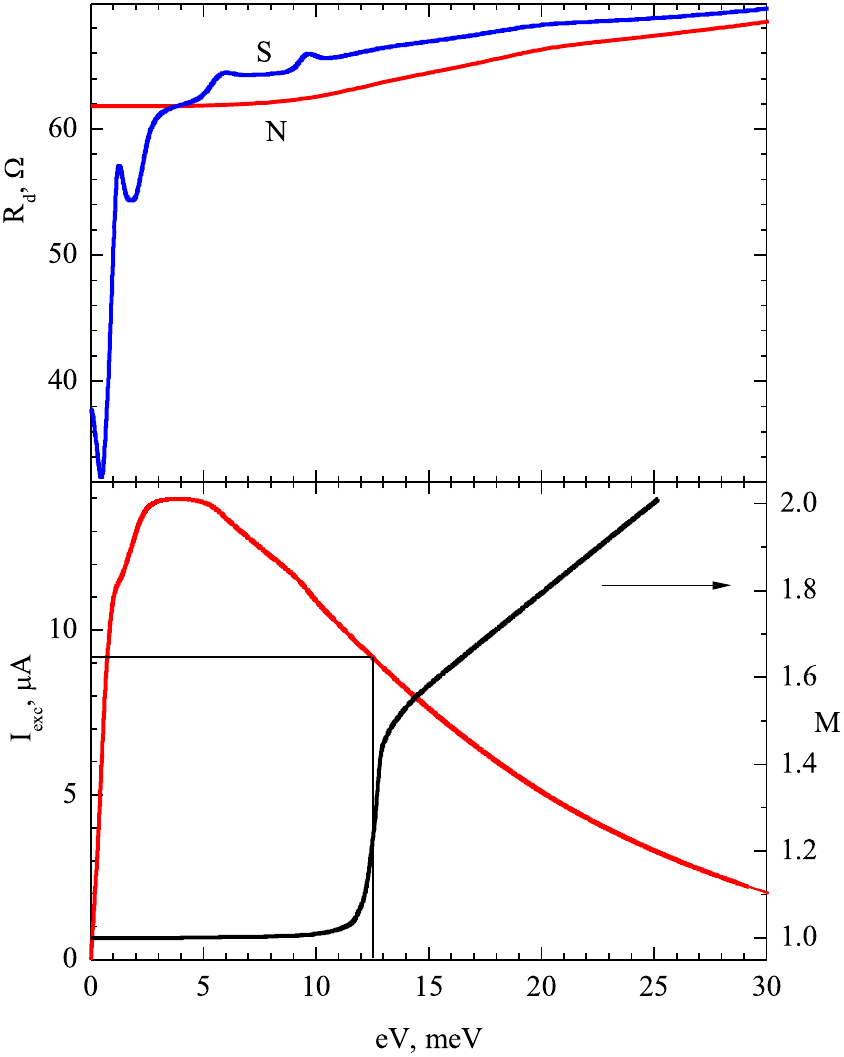}
\caption[]{Differential resistances of the Ta-Ta point contact whose characteristics
are shown in Fig.\ref{Fig11} in the normal (N) and superconducting (S) states and
excess current vs. the bias. The excess current panel shows the scaling curve
M (see Fig.\ref{Fig11}) with designated excess current and voltage values corresponding
to the start of the ascending part of the scaling curve.}
\label{Fig12}
\end{figure}

Figure \ref{Fig12} shows the differential resistances of the point
contact in the normal and superconducting states, the energy
dependence of the excess current, and the scaling curve. As can
be seen in Fig.\hyperref[Fig11]{11(d)}, the sharpening and the intensity growth of
the first peak upon the superconducting transition are markedly
less pronounced in this case than for heterocontacts with a comparable
resistance. Regarding the superconducting contribution to
the spectrum [Fig.\hyperref[Fig11]{11(b)}], its high-energy part is comparable in
intensity to the low-energy one. Therefore, the correction scaling
curve M is relatively small in amplitude as compared with that of
the heterocontacts.
\begin{figure}[tbp]
\includegraphics[width=8cm,angle=0]{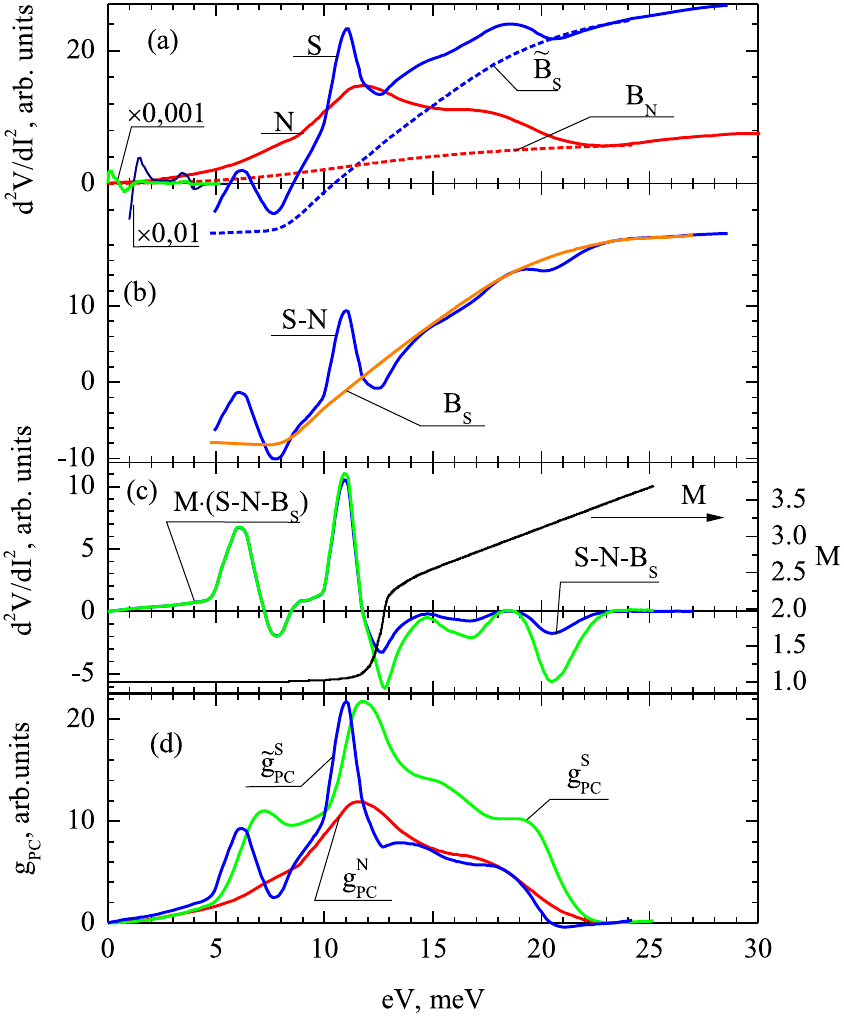}
\caption[]{a) EPI spectra of the Ta-Ta point contact in the normal and superconducting
states; N: $T=4.6 K$; S: $T=2.55 K$, $R_0$=17~ohm; the initial segments of
the superconducting curve corresponding to the critical current and gap and
nonequilibrium effects are scaled down by factors of 1000 and 100, respectively,
$\tilde{B}_S$ and $B_N$ are the background curves for the normal and superconducting
spectra. (b) The difference between the superconducting and normal spectra
and the assumed shape of the background curve. (c) Difference curve (after
background subtraction), scaling curve M, and the difference curve multiplied by
the scaling curve. (d) Point contact EPI functions recovered from the spectra for
normal and superconducting states and from the superconducting contribution
to the spectrum by integrating the corrected difference curve (c). For the convenience
of comparison, the curve $g^S_{pc}$ is aligned in amplitude with the curve $\tilde{g}^S_pc$.
The scale is the same in all panels of the Figure.}
\label{Fig13}
\end{figure}

\begin{figure}[tbp]
\includegraphics[width=8cm,angle=0]{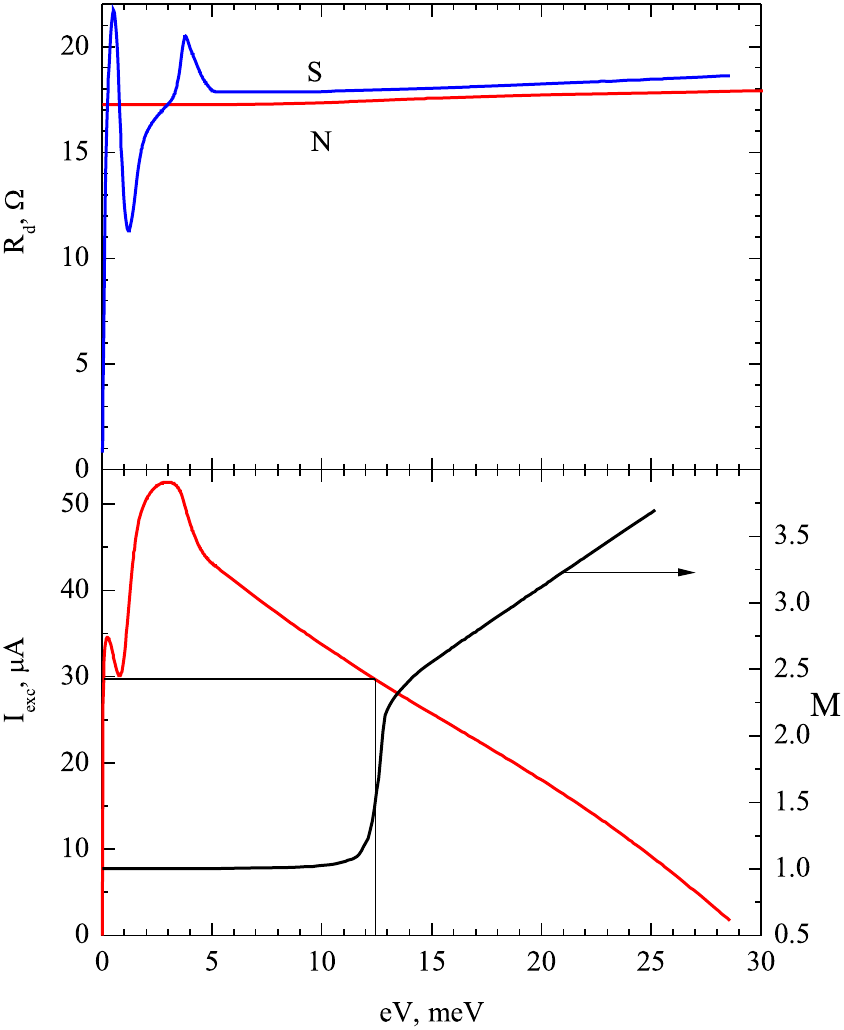}
\caption[]{Differential resistance of the Ta-Ta point contact whose characteristics
are shown in Fig.\ref{Fig13} in the normal (N) and superconducting (S) states and
excess current vs. the bias. The excess current panel shows the scaling curve
M (see Fig.\ref{Fig13}) with designated excess current and voltage values corresponding
to the start of the ascending part of the scaling curve.}
\label{Fig14}
\end{figure}
Characteristics for the markedly lower-resistance homocontact
(17~ohm) are shown in Figs. \ref{Fig13} and \ref{Fig14}. For this contact, the
nonequilibrium effect was found to occur at less than 5~mV
energy; therefore, the transformation of the soft mode during the
superconducting transition of the contact was accessible for
observation. Like for heterocontacts, it is converted into a peak.
As follows from Fig.\hyperref[Fig13]{13(d)}, pronounced sharpening and an
amplitude increase of the first peak are also observed for this
contact. Regarding the high-energy part of the spectrum, the
changes induced by the superconducting transition are moderate.
Nevertheless, a comparison with the proportion of the
high-energy part of the superconducting contribution for the
heterocontact with virtually the same resistance [Fig.\hyperref[Fig8]{8(c)}]
demonstrates that this proportion is quite comparable in intensity
with that of the low-energy part; therefore, the correction
scaling curve is an order of magnitude smaller.

\section{RESULTS AND DISCUSSION}
\begin{figure}[tbp]
\includegraphics[width=8cm,angle=0]{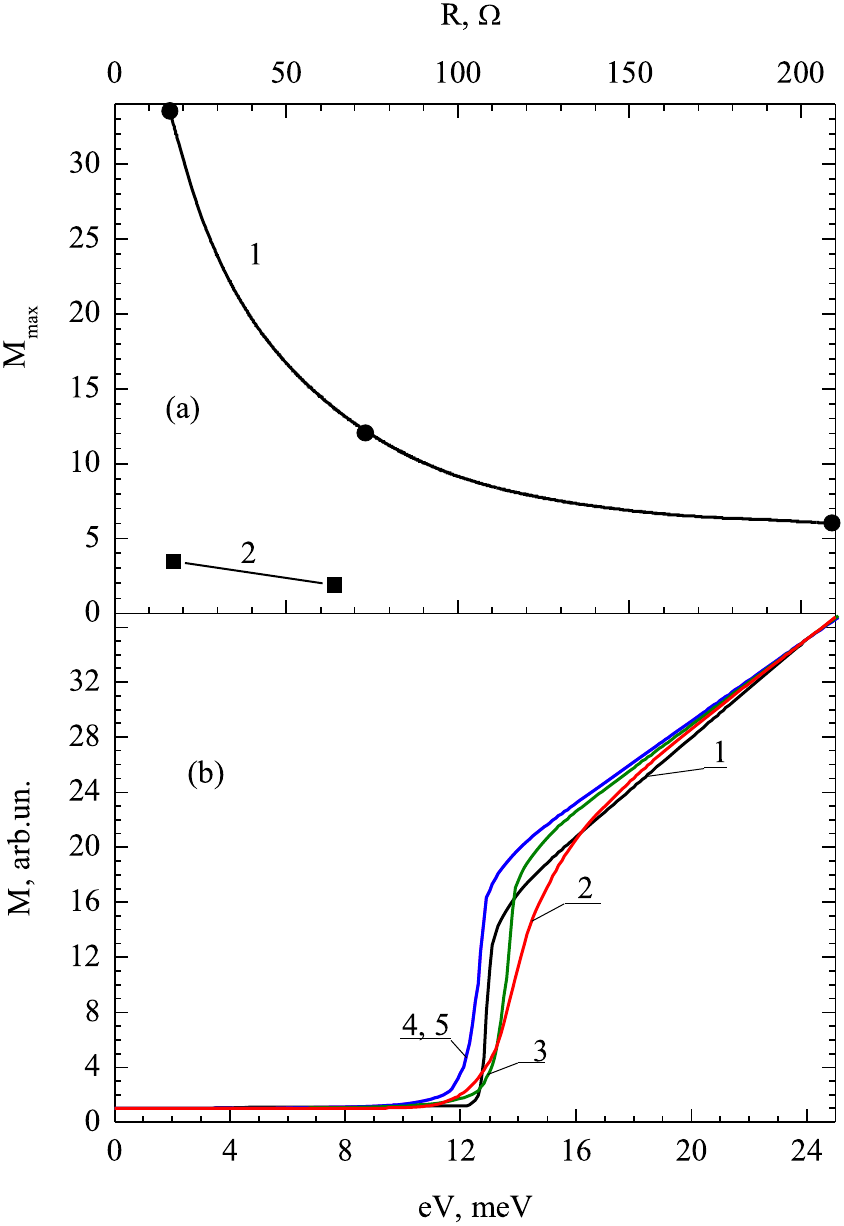}
\caption[]{(a) A correction scaling curve at the phonon spectrum boundary at 23~meV as a function of resistance of hetero- and homocontacts. (b) Correction
scaling curves for contacts with different resistances reduced to the same amplitude;
(1) Ta-Cu, $R$=209~ohm; (2) Ta-Cu, $R$ = 73~ohm; (3) Ta-Cu, $R$ = 16~ohm;
(4) Ta-Ta, $R$ = 64~ohm; (5) Ta-Ta, $R$ = 17~ohm. Curves \emph{4} and \emph{5} have the same
shape but differ in the amplitude approximately 3.7-fold (see Figs.\ref{Fig12} and \ref{Fig14}).}
\label{Fig15}
\end{figure}

The conducted study of the ballistic point contacts based on
tantalum demonstrated that all deviations from the predictions of
the theoretical model are caused by involving a region that is about
the size of the coherence length, which is adjacent to the constriction,
in the formation of the EPI spectra upon the transition of the
contact to the superconducting state. Apart from other differences,
these contacts show a pronounced dependence of the superconducting
contribution to the spectrum on the contact bias and on the
resistance. This dependence, caused by partial suppression of the
excess current by nonequilibrium quasi-particles, complicates the
recovery of the EPI function from this contribution and requires a
correction of its shape to ensure the fulfilment of the rule of sums,
after subtraction of the superconducting background.

The search for the shape and intensity of the correction scaling
curve was performed separately for each of the considered point
contacts. As has already been noted, the scaling curve intensity
varies over wide limits, which is reflected in Fig.\hyperref[Fig15]{15(a)}. However, the
shapes of these curves were similar. Figure \hyperref[Fig15]{15(b)} shows all scaling
curves M reduced to the same amplitude. Moreover, the same correction
curve proved to be applicable, with only the amplitude being
varied. For example, the curves for contacts \emph{4} and \emph{5} have the same
shape, but differ in the amplitude approximately 3.7-fold. Attention
is attracted by the step present in all correction curves at similar
energies with approximately the same amplitude on a reduced scale.
The curves for the excess current as a function of the bias show no
peculiar effects at these energies, with the magnitude of the excess
current smoothly descending. By analogy with appearance of a step
along the dependence of the excess current on the bias, associated
with the attainment of the critical concentration of nonequilibrium
quasi particles above the gap in the near-contact region, one can
assume that there exists some threshold concentration of Andreev
electrons determined by the excess current, below which the
efficiency of reabsorption of nonequilibrium phonons sharply
decreases, resulting in a sharp decrease in the superconducting contribution
to the spectrum. Further decreases in the concentration of
Andreev electrons with increasing contact bias lead to a smooth
decline of the superconducting contribution to the spectrum, which
is reflected in the shape of the correction scaling curves. This
assumption is supported by the comparison of homo- and heterocontacts.
When the resistances are similar, the amplitude of the correction
curves is much smaller for homocontacts, which is due to
the doubled excess current. However, as can be seen in Fig.\ref{Fig15},
there are no obvious differences between the shapes of the correction
curves for homo- and heterocontacts.

\section{BRIEF CONCLUSIONS}
\begin{itemize}
\item [1.]{The tantalum-based ballistic point contacts in the superconducting
state show deviations from the predictions of theoretical
models due to the involvement of a near-contact region with a
size that is about equal to the reduced coherence length $\zeta$, in the
formation of EPI spectra.}
\item [2.]{The relative value of the superconducting contribution to the
EPI spectrum depends on the contact resistance and increases
with increasing contact diameter.}
\item [3.]The superconducting contribution decreases with increasing
contact bias due to the suppression of excess current by the
nonequilibrium quasi-particles, which leads to violation of the
formulated rule of sums.
\item [4.]Although the EPI functions are proportional to the first derivative
of the excess current, the presence of the superconducting
background and the need to correct the amplitude of the superconducting
spectral contribution caused by violation of the rule of sums, bring about the necessity to recover the EPI functions
from the CVC second derivatives.
\item [5.]The procedure of recovering the EPI spectral function described
in detail can be used to analyze the characteristics of ballistic
point contacts based on a broad range of superconductors.
\end{itemize}

\section{ACKNOWLEDGMENTS}
This study was supported by the National Academy of
Sciences of Ukraine as a part of project FTs 4-19.

In conclusion, the author would like to express gratitude to
A.V. Khotkevich for numerous tips, comments, and discussions.


\begin{thebibliography}{}
\bibitem{1}I.K. Yanson and A. V. Khotkevich, Atlas Mikrokontaktnykh Spektrov Elektron-
Fononnogo Vzaimodeistviya v Metallakh (Naukova Dumka, Kyiv, 1986)
[A. V. Khotkevich and I. K. Yanson, \href{https://link.springer.com/chapter/10.1007%2F978-1-4615-2265-2_2}{Atlas of Point-Contact Spectra of Electron-
Phonon Interaction in Metals}, (Kluwer Academic Publishers, Boston, 1995)].
\bibitem{2}Y. G. Naidyuk and I. K. Yanson, \href{https://link.springer.com/book/10.1007%2F978-1-4757-6205-1}{Point-Contact Spectroscopy} (Springer, New York,
2005).
\bibitem{3}I. K. Yanson, I. O. Kulik, A. N. Omel'yanchuk, R. I. Shekhter, and
Y. V. Sharvin, Phenomenon of Charge Carrier Energy Redistribution in Metal
Microcontacts at low Temperatures (Discoveries in the USSR, VNIIPI, Moscow,
1986), p. 18 (Diploma No. 328, Discoveries. Invent. (1987), No. 40, p. 3).
\bibitem{4}I. O. Kulik, A. N. Omel'yanchuk, and R. I. Shekhter, \href{https://fntr.ilt.kharkov.ua/fnt/pdf/3/3-12/f03-1543r.pdf}{Fiz. Nizk. Temp.} 3, 1543
(1977) [Sov. J. Low Temp. Phys. 3, 740 (1977)].
\bibitem{5}V. A. Khlus and A. N. Omel'yanchuk, \href{https://fntr.ilt.kharkov.ua/fnt/pdf/9/9-4/f09-0373r.pdf}{Fiz. Nizk. Temp.} 9, 373 (1983) [Sov.
J. Low Temp. Phys. 9, 189 (1983)].
\bibitem{6}V. A. Khlus, \href{https://fntr.ilt.kharkov.ua/fnt/pdf/9/9-9/f09-0985r.pdf}{Fiz. Nizk. Temp.} 9, 985 (1983) [Sov. J. Low Temp. Phys. 9, 510,(1983)].
\bibitem{7}N. L. Bobrov, V. V. Fisun, O. E. Kvitnitskaya, V. N. Chenobai, and I. K. Yanson, \href{https://fnt.ilt.kharkov.ua/fnt/pdf/38/38-5/f38-0480r.pdf}{Fiz. Nizk. Temp.} 38, 480 (2012) [\href{https://doi.org/10.1063/1.4709437}{Low Temp. Phys.} 38, 373 (2012)]; \href{https://arxiv.org/pdf/1207.6486.pdf}{arXiv:1207.6486}.
\bibitem{8}N. L. Bobrov, A. V. Hotkevich, G. V. Kamarchuk, and P. N. Chubov, \href{https://fnt.ilt.kharkov.ua/fnt/pdf/40/40-3/f40-0280r.pdf}{Fiz. Nizk.Temp.} 40, 280 (2014) [\href{https://doi.org/10.1063/1.4869565}{Low Temp. Phys. }40, 215 (2014)]; \href{https://arxiv.org/pdf/1405.6869.pdf}{arXiv:1405.6869}.
\bibitem{9}N. L. Bobrov, \href{https://fnt.ilt.kharkov.ua/fnt/pdf/41/41-8/f41-0768r.pdf}{Fiz. Nizk. Temp.} 41, 768 (2015) [\href{https://doi.org/10.1063/1.4929594}{Low Temp. Phys.} 41, 595 (2015)]; \href{https://arxiv.org/pdf/1512.04082.pdf}{arXiv:1512.04082}.
\bibitem{10}I. K. Yanson, V. V. Fisun, N. L. Bobrov, and L. F. Rybal'chenko, \href{http://www.jetpletters.ac.ru/ps/141/article_2443.shtml}{JETP Letters}
45, 425 (1987) [\href{http://www.jetpletters.ac.ru/ps/1244/article_18813.shtml}{JETP Lett.} 45, 543 (1987)]; \href{https://arxiv.org/pdf/1602.04356.pdf}{arXiv:1602.04356}.
\bibitem{11}I. K. Yanson, N. L. Bobrov, L. F. Rybal'chenko, and V. V. Fisun, \href{https://fntr.ilt.kharkov.ua/fnt/pdf/13/13-11/f13-1123r.pdf}{Fiz. Nizk.
Temp.} 13, 1123 (1987) [Sov. J. Low Temp. Phys. 13, 635 (1987)]; \href{https://arxiv.org/pdf/1512.03917.pdf}{arXiv:1512.03917}.
\bibitem{12}I. K. Yanson, L. F. Rybal'chenko, N. L. Bobrov, and V. V. Fisun, \href{https://fntr.ilt.kharkov.ua/fnt/pdf/12/12-5/f12-0552r.pdf}{Fiz. Nizk. Temp.} 12, 552 (1986) [Sov. J. Low Temp. Phys. 12, 313 (1986)]; \href{https://arxiv.org/pdf/1512.00684.pdf}{arXiv:1512.00684}.
\bibitem{13}N. L. Bobrov, L. F. Rybal'chenko, V. V. Fisun, and I. K. Yanson, \href{https://fnt.ilt.kharkov.ua/join.php?fn=/fnt/pdf/13/13-6/f13-0611r.pdf}{Fiz. Nizk. Temp.} 13, 611 (1987) [Sov. J. Low Temp. Phys. 13, 344 (1987)]; \href{https://arxiv.org/pdf/1512.01800.pdf}{arXiv:1512.01800}.


\end{thebibliography}
\end{document}